\documentclass[11pt, preprint]{aastex}
\usepackage{psfig}
\def\edcomment#1{\iffalse\marginpar{\raggedright\sl#1\/}\else\relax\fi}
\reversemarginpar

\begin{document}
\title{SURFACE COMPOSITIONS OF R CORONAE BOREALIS STARS \& EXTREME HELIUM STARS
        - SOME CONNECTIONS}
\author{N.Kameswara Rao}
\affil{Indian Institute of Astrophysics, Bangalore -560034,
India}

\begin{abstract}
 Abundances of R Coronae Borealis stars (RCBs) and 
  Extreme Helium stars (EHes) are
   discussed. Recent estimates of the $s$-process elements in hot extreme
   helium stars show enhancements of lighter $s$-process elements (Y, Zr)
   relative to heavier $s$-process elements, a characteristic shared by RCB
   stars. It also suggests that atleast some EHe stars went through an episode
   of $s$-process element synthesis in their earlier evolution. A majority of
   RCB stars show a high $^{12}$C/$^{13}$C ratio in their atmospheres. 
   A recent analysis
   of the spectrum of minority RCB star V\,CrA, however, shows a lower value
   between 4 $-$ 10. The implications  of these results are discussed.
\end{abstract}
\thispagestyle{plain}

\section{Introduction}

         It is a great pleasure to be able to participate in this celebration   honouring Prof. David Lambert. I would like to convey to David greetings and   best wishes from his numerous friends and admirers in India. It is almost a
   hundred years since Ludendorff discovered the absence of Balmer lines in the  spectrum of the F type irregular variable R\,CrB, the first recognised hydrogen
deficient star. About 35
   years later, the first helium star was  discovered by Popper.  The basic
   problem about the origins of these stars, namely,  a single (or double)  
    star of intermediate mass becoming  a high luminosity hydrogen deficient star,
   still remains  a mystery. Presently, there are two principal hypotheses
   to account for their origins. The prime contender is the white dwarf merger
   scenario (or double degenerate, DD) in which a helium white dwarf secondary
   is accreted onto a CO (or He) white dwarf primary resulting in the
   ignition of a helium shell in the accreted envelope. The energy generated
   forces the star to expand to cool supergiant dimensions. The life time at
   this stage depends on the supply of helium in the shell and the mass of the
   envelope. The subsequent evolution would be like a canonical post-AGB
   contraction to the white dwarf track (Webbink 1984, Iben \& Tutukov 1984,
   Saio \& Jeffery 2000).
   The second scenario proposed is called the late thermal pulse (or final
   flash, FF), in which an AGB star in its final stages retains a
   helium shell of enough mass  as to get ignited for a last time after it
   descends on to the white dwarf track. The final thermal pulse forces the
   star to become a cool supergiant for a second time. The subsequent evolution 
   would again resemble the canonical post-AGB evolution of single stars (Iben et al.
   1983).
   Presently, there are no decisive observational tests predicted to choose
   either one (or any other alternative) of the scenarios. 
   
   Several groups of
   hydrogen deficient
  stars have now been identified ranging from high luminosity (low surface
  gravity) cooler HdC stars, RCBs, EHes to low luminosity (high gravity)
  hot PG\,1159 stars (see Jeffery 1996 for a display of various groups in the
log $g$, log $T_{\rm eff}$ plane). It is not clear how these various groups of
  hydrogen deficient stars are related to each other $-$ e.g., EHes and RCBs. Do
they
  represent sucessive stages in evolution? Even some groups are further
  subdivided and show diversity in properties e.g., majority and minority RCBs
  which show [S/Fe], [Si/Fe] values of about 0.5 and much higher, respectively
  (Lambert \& Rao 1994, Asplund et al. 2000). Do they suggest different paths
  of evolution? In the last few years, stars like Sakurai's object and FG\,Sge
  provided us examples of stellar evolution in real time by turning from a
  normal star to hydrogen deficient born-again giant (Asplund et al. 1997, Gonzalez
  et al. 1998, Duerbec et al. 2002). How are these born-again giants related
  to RCBs are some of the intriguing questions that the theory of intermediate
  mass star evolution is unable to provide. In keeping with the theme of the
  conference,`abundances as records of stellar evolution and nucleo-
  synthesis', I would like to explore in this presentation the
  interrelationships of RCBs and EHes and also address some of the issues raised above
  by seeking clues from surface chemical compositions.

\section{Properties}

     R Coronae Borealis (RCB) stars are carbon-rich, hydrogen-poor F-G supergiants
  that exhibit light fades of several magnitudes at irregular intervals due
  to circumstellar dust formation. They range in $T_{\rm eff}$ from 8000 to 4000 K.
  After Keenan \& Barnbaum's (1997) spectroscopic investigation even the cool
  so-called DY\,Per stars also seem to belong to this class, extending the
  $T_{\rm eff}$ to 3500 K. The effective gravities range from log $g$ = 0.5 to 1.5.
  The RCB population in the
  LMC provides the estimates of absolute magnitudes which range from $M_{V}$
  $-$2.5 to $-$5 and corresponding luminosities log (L/$L_{\odot}$) of 4.0 $-$ 3.2
  (Alcock et al. 2001).  Presently known members in the Galaxy amount to about
  35 including the 3 hot RCBs, about 21 are known in the LMC (including DY Per
  stars) and a lone one in the SMC (Alcock et al. 2001, Morgan et al. 2003).
  The distribution of the number of RCBs with respect to spectral type suggests
  a peak around F-G for Galactic ones whereas the LMC population shows a peak
  at much cooler temperatures. By scaling the LMC population of RCBs to the
  Galaxy, Alcock et al. estimate more than 3200 to be present in the galaxy
  most of them being cooler members. Any evolutionary schemes that are proposed
  to explain the origins
  should also be able to account for these numbers.
  The two evolutionary schemes that have been  proposed (DD and FF)
  can account only partially for the estimated number of RCBs present in the Galaxy
  (Iben, Tutukov, Yungelson 1996).
       Extreme helium stars are mostly carbon rich, hydrogen poor
  A-B supergiants. Some of them show short period pulsations with a period of a
  few days. They range in $T_{\rm eff}$ from 32000 to 9000 K with log $g$ of 0.7 to 4.0
  . The luminosities, log L/$L_{\odot}$ ,are estimated to be
  about  4.4 to 3.0 (Saio \& Jeffery 2000 ). Presently there are
  about 21 known members in the Galaxy and none in either LMC or SMC.

     The Galactic distribution of both groups (RCB and EHe) suggests a bulge population
  (Jeffery et al. 1987) and might belong to the thick disk, although there are
  suggestions that a few of them might even be part of the halo (eg. U Aqr -
  Cottrell \& Lawson 1998).

      The first major study of the surface abundances of a larger number of
  RCBs by Lambert \& Rao (1994) revealed that majority of the stars analysed
  (14 out of 18) showed similar patterns; particularly [Si/Fe] and 
  [S/Fe]\footnote{the [$i$] notation refers to log $\epsilon$(i)$^{\star}$ $-$ 
log $\epsilon$(i)$^{\odot}$} are
  around 0.5 and a mild Fe deficiency relative to solar. The minority
  RCBs are marked by approximately solar Si and S abundances and a severe Fe deficiency
  (or a very high [Si/Fe], [S/Fe]). A similar classification  by  Fe abundance
was also
  suggested for EHe stars (Heber 1986).

\section{Spectral analyses}

      Before discussing the surface abundances it is appropriate to recall
  some of the uncertainities involved in arriving at these estimates.
  Most of the analyses are based on the line-blanketed, LTE atmospheric
  models computed at Uppsala by Asplund et al. (1997) for RCB stars of $T_{\rm eff}$
  8000 to 6000 K and at Armagh by Jeffery \& Heber (1992), Jeffery, Woolf, Pollacco (2001) for EHes with $T_{\rm eff}$ $>$ 9000 K.
  The continuous opacity in the atmospheres of RCBs is controlled by the
  photoionization of C\,{\sc i} from excited levels and the gas pressure is provided by
  helium. In estimating the mass fraction of elements, a crucial parameter
  needed for RCBs is the C/He ratio (the number density of carbon to helium), which
  can not be estimated from the spectrum directly (Rao \& Lambert 1996,
  Asplund et al. 2000). This ratio in EHes, which can be directly estimated 
  from the analysis of  spectral lines (Jeffrey 1996, Pandey et al. 2001)
  has a  mean value of 1\%. It has been assumed in the analyses of RCBs that
  the same C/He of 1\% holds. This value of  C/He seems to be consistent with the
  metallicity expected from galactic distribution (Rao \& Lambert 1996, Pandey et al. 2001).
        One of the surprising outcomes of the atmospheric analysis of RCBs is
  the so called `carbon problem', as discussed in detail by Asplund et al.
  (2000)(first noticed by David Lambert). The carbon abundance estimated from
  the observed C\,{\sc i} lines is four times less than the input carbon abundance
  for C/He, with less input the C/He ratio is very very small.  Recently, based on the analysis of [C\,{\sc i}] lines 9850\AA\ and 8727\AA,
  Pandey et al. (2004b) suggest that a chromosphere like temperature rise in the
 atmosphere might be able to account for the carbon problem. However it was
  realised by Asplund et al. (2000) that the abundance ratios are largely unaffected by the
  carbon problem (and the assumed C/He values). An $M_{bol}$ of $-$5 has been assumed for all RCBs
  in arriving at the appropriate $T_{\rm eff}$ and log $g$, although it is known that LMC
  RCBs show a range of values. As discussed by Pandey et al. (2001), the 
  element which represents the initial metallicity of the stars is not clear, whether
  it is Fe or Si, S. A metallicity parameter M has been defined by Pandey et al.
  based on the Si, S abundances that represents initial Fe abundance. The
  observed Fe abundance might be affected by things other than the intial
  metallicity of the star. This property might even apply to hot RCB stars like
  DY  Cen as well. The extreme Fe deficiency is suggested to be a result of the
  accretion of winnowed gas from dust (Jeffery \& Heber 1993).

\section{Abundance Patterns}

       The surface abundances of RCBs and EHes have been estimated by Asplund
  et al. (2000), Rao \& Lambert (2003, 2004), Jeffery (1996), Pandey et al.
  (2001) and Pandey et al. (2004a). Both RCBs and EHes show the majority,
  minority division proposed by Lambert \& Rao (1994). Mainly they are
  differentiated by Fe abundance. The majority cluster around [Fe] of $-$1 and
  the minority show a larger deficiency about $-$1.7 or more. Since a C/He of 1\%
  is assumed for all stars (except the minority RCB star V854\,Cen for which 
  a value of 10\% is
  suggested -Asplund et al. 1998), the absolute numbers can be compared.
  15 stars out of 19 analysed RCBs comprise the majority and 12 out of 14 EHes
  analysed constitute the majority class. The 4 minority RCBs are V\,CrA, VZ\,Sgr,
  V3795\,Sgr and V854\,Cen and the 2 minority EHes are BD +10 2179 and FQ\,Aqr.
  The mean abundances (normalised to $\log\Sigma\mu_X\epsilon$(X) = 12.15
  with $\mu$ as the atomic weight) of each group are shown in Table 1. 
  The dispersion around

\begin{table}[!ht]
\caption{Abundances}
\smallskip
\begin{center}
{\small
\begin{tabular}{llllll}
\tableline
\noalign{\smallskip}
        &       & \multicolumn{4}{c}{ }\\
Element & Z     & Maj. RCBs(15)  & Min. RCBs(4) & Maj. EHes(12) & Min. EHes(2)\\
\noalign{\smallskip}
\cline{3-6}
\noalign{\smallskip}
\tableline
\noalign{\smallskip}
H    &  1  & 6.14$\pm$0.89  & 7.57 & 7.19$\pm$0.95 & 7.35 \\
He   &  2  & 11.54          & 11.54  & 11.54  & 11.54 \\
C    &  6  & 8.91$\pm$0.14  & 9.0$\pm$0.35 & 9.32$\pm$0.22 & 9.05$\pm$0.05\\
N    &  7  & 8.67$\pm$0.23  & 7.88$\pm$0.20 & 8.36$\pm$0.34 & 7.63$\pm$0.48\\
O    &  8  & 8.17$\pm$0.41  & 8.22$\pm$0.59 & 8.60$\pm$0.48 & 8.5$\pm$0.4\\
Ne   & 10  & 8.3(1)         & 7.9(1)        & 9.11(6)$\pm$0.25 & 7.9(1) \\
Na   & 11  & 6.13$\pm$0.22  & 5.94          & 6.5(3)$\pm$0.8 & \\
Mg   & 12  & 6.72(5)$\pm$0.21 & 6.3(3)$\pm$0.22 & 7.40$\pm$0.35 & 6.5$\pm$0.5 \\
Al   & 13  & 5.95$\pm$0.29    & 5.51$\pm$0.14 & 6.07$\pm$0.52 & 5.5$\pm$0.8 \\
Si   & 14  & 7.12$\pm$0.19    & 7.34          & 7.23$\pm$0.60 & 6.5$\pm$0.2 \\
P    & 15  & 5.9(1)           & 6.5(1)        & 5.62$\pm$0.47 & 4.85$\pm$0.65 \\
S    & 16  & 6.87$\pm$0.33    & 6.93$\pm$0.40 & 7.08$\pm$0.32 & 6.55$\pm$0.55 \\
Ca   & 20  & 5.36$\pm$0.19    & 5.16$\pm$0.07 & 5.84(5)$\pm$0.27 & 4.2(1) \\
Sc   & 21  & 2.87(6)$\pm$0.19 & 2.89          & 3.3(1)           & 2.1(1) \\
Ti   & 22  & 4.03(8)$\pm$0.14 & 3.6(3)$\pm$0.34 & 4.53(3)$\pm$0.17 & 3.25$\pm$0.05 \\
Fe   & 26  & 6.49$\pm$0.24    & 5.73          & 6.89$\pm$0.32    & 5.7$\pm$0.3 \\
Ni   & 28  & 5.82$\pm$0.24    & 5.44$\pm$0.43 & 5.93(3)$\pm$0.47 & 5.0(1) \\
Zn   & 30  & 4.34(13)$\pm$0.28 & 4.08$\pm$0.21 & 4.4(2)$\pm$0.2 & 4.14(1) \\
Y    & 39  & 2.08$\pm$0.50     & 1.94$\pm$0.60 & 2.27$\pm$0.59  & 1.75(1) \\
Zr   & 40  & 2.09(8)$\pm$0.29  & 2.04$\pm$0.48 & 2.7(4)$\pm$0.60 & 1.83$\pm$0.53 \\
Ba   & 56  & 1.43$\pm$0.56     & 0.97$\pm$0.43 & 1.7      & 0.5 \\
\noalign{\smallskip}
\tableline
\end{tabular}
}
\end{center}
\end{table}

  the mean in each group is surprisingly small particularly for the majority
  groups of RCBs and EHes $\sim$ 0.27 dex. Only H and the $s$-process elements show a
  little more dispersion. The similarity in the abundance
  pattern of the majority RCBs and EHe (Fe of 6.5 and 6.8, respectively) is striking.
  The mean difference for 15 elements is 0.23 dex. However H, N, Ne and Mg
  show significant differences. The minority groups RCBs and EHes (Fe of 5.7
  for both) also show small differences ($<$ 0.13 dex) for most elements except
  Si, Ca (may be P and Sc).
         The fact that the H abundance for the majority of RCBs is lower by 1.0 dex 
  compared to
  EHes and the N abundance is also higher by 0.3 dex might suggest that
  RCBs are a later phase in evolution to EHes, However, the larger abundance of
  Ne and Mg in EHes furthur indicates that $^{14}$N is converted to $^{22}$Ne and
  $^{25}$Mg by alpha processing, thus EHes might be a later phase to RCBs
  as is expected from the tracks of post AGB stars in the log $g$, log $T_{\rm eff}$ plane.

\begin{figure}[ht!]
\psfig{figure=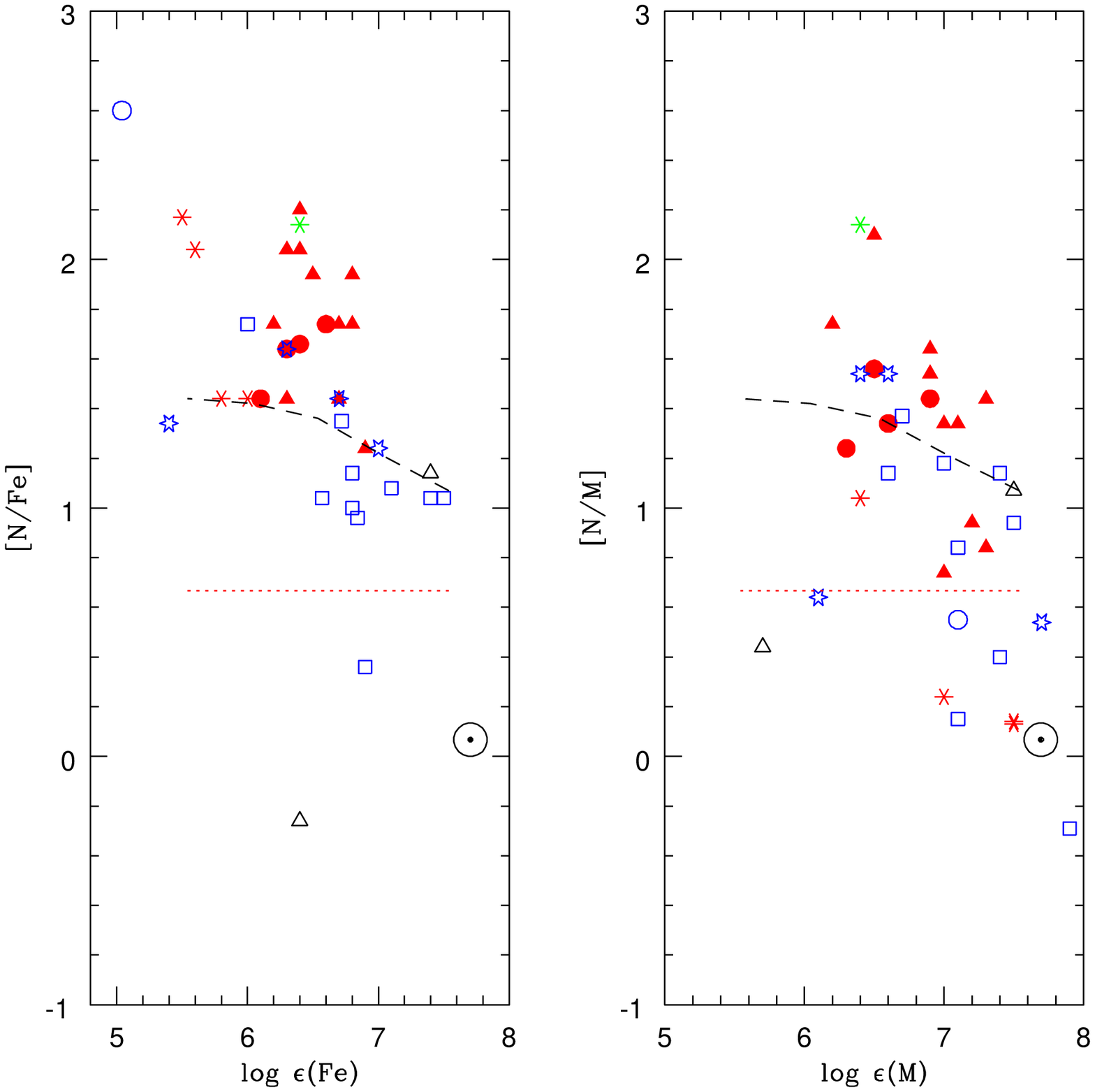,width=\hsize,height=14cm}
\caption{Nitrogen abundance in RCBs and EHes. Solid triangles and dots refer to majority
RCBs, asterisks refer to minority RCBs, stars refer to cool EHes, open squares hot EHes, circle refers to DY Cen.}
\end{figure}

\subsection{CNO abundances}

 Figure 1 shows [N/Fe] versus Fe and [N/M]versus
  M (the metalicity parameter) for both groups of RCBs and EHes. Clearly, most of the RCBs have N abundances
  that are predicted from conversion of initial C and O to N or even more. In
  some cases C, produced in the He burning, might also have been converted to N.
  The newly discovered RCB star V2552\,Oph may be one such and illustrates the N
  enhancement prominently (Rao \& Lambert 2003).

   Although R\,CrB and V2552\,Oph have very
  similar line spectra and physical parameters, V2552 Oph shows much stronger N\,{\sc i} lines
 than R CrB.  On the other hand, the N abundance in EHes
  generally lies between the expected N from conversion of initial C to N and
  the value of N expected from conversion of initial C, O both to
  N . The minority stars (both groups and DY\,Cen) have N abundances
  less than that expected from full convertion of C to N.
      Although the N abundance in many RCBs and in some EHes imply wholesale
  conversion of O to N via ON cycles, many stars are not O deficient suggesting
  O is synthesized along with C, i.e., 3$\alpha$ -process was followed by
  $^{12}$C$(\alpha,\gamma)^{16}$O. Most of the O-rich stars have an observed O/C about 1 implying
  equal production of C and O.
  
 \begin{figure}[ht!]
\psfig{figure=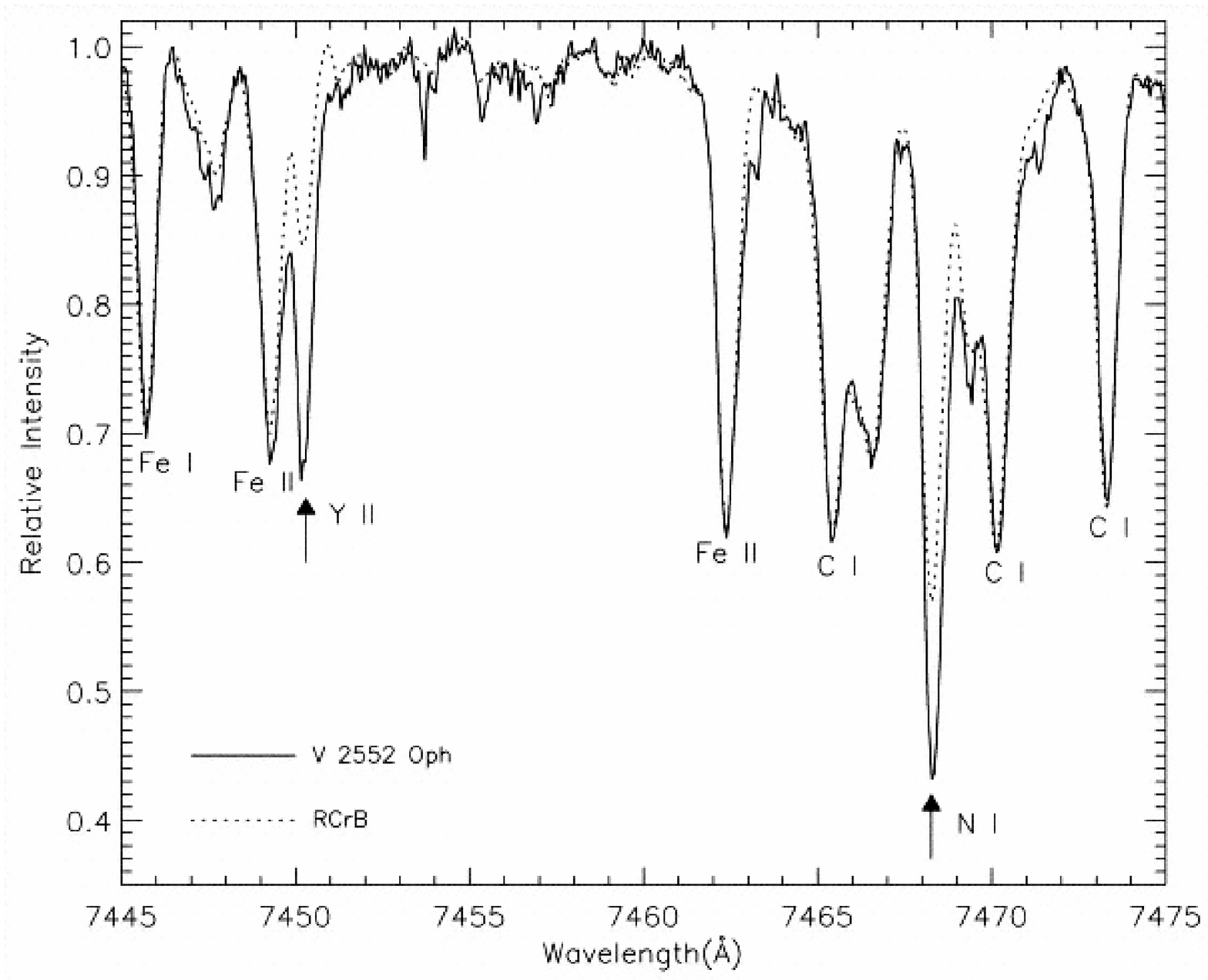,width=\hsize,height=10cm}
\caption{Enhanced lines of N I, and Y II in V2552 Oph relative to R CrB.}
\end{figure}
  
\subsection{$s$-Process elements}

Dramatic enhancement of light $s$-process elements
  Sr, Y, and Zr was first seen in cool RCB star U\,Aqr (Bond et al. 1979).
  Asplund et al. (2000) have shown that most of the RCB stars show enhancements
  [Y/Fe] of about 0.8 and [Ba/Fe] of about 0.4, i.e., the lighter $s$-process elements
  are more enhanced than the heavier ones. However, there is a
  considerable dispersion in the Y and Ba  abundances in RCBs. 
  
  Both Y and Ba abundances
  seem to be anticorrelated with H abundance. U\,Aqr shows extraordinary overabundances of 
  $s$-process elements [Y/Fe] $\sim$ 3.3 and [Ba/Fe] $\sim$ 2.1
  (Vanture et al. 1999). Generally these enhancements relative to Fe are
  consistent with a mild single neutron exposure $\tau_{o}$ $\sim$ 0.1 mb$^{-1}$
  These estimates of $s$-process abundances are 
  not available for
  EHes stars (Asplund et al. 2000).  Did  EHe stars pass through a phase of $s$-process element
  production (similar to RCBs ) ? Did they undergo  third dredgeup
  and show $s$-process
  abundance pattern similar to AGB or post -AGB stars. These are some of the
  questions that need to be addressed.
  
   \begin{figure}[ht!]
\psfig{figure=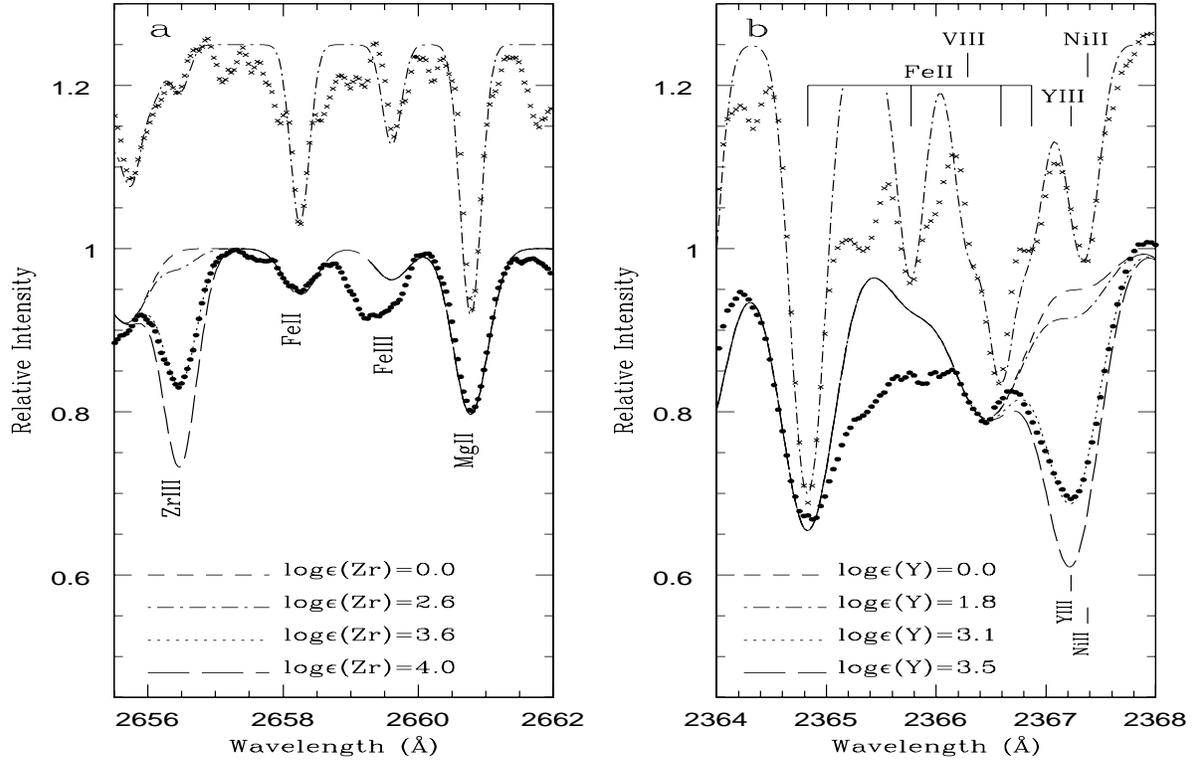,width=\hsize,height=11cm}
\caption{Enhanced lines of Y III, Zr III in V1920 Cyg (lower spectrum
            - dots) relative to HD124448}
\end{figure}

  In the atmospheres of the hot EHes most of the $s$-process elements exist in
  doubly ionized state and lack spectral lines in the optical region.
  [Let me add a personal note here. In 1996 January, David Lambert and I were attending a
  conference on spectroscopy in Bombay where we heard a talk by Indrek Martinson
  discussing  the Zr\,{\sc iii} and Y\,{\sc iii} spectra in UV and the availability of fairly
  decent $gf$-values. This prompted us to apply for $HST$ - $STIS$ spectra in
  search of Zr and Y abundances in
   EHe stars.] Fortunately, strong lines of Y{\sc iii}, Zr\,{\sc iii}, Ce\,{\sc iii}, La\,{\sc iii} etc.,
  do occur in the UV where EHe stars have appreciable flux. We could obtain
  UV spectra of 7 EHe stars with $STIS$ on $HST$. Analyses of the spectra of two
  EHe stars V1920\,Cyg and HD\,124448 demonstrate the similarities in the
  pattern of $s$-process elements with RCBs. The two stars have the same log $T_{\rm eff}$
  and log $g$ but show large differences in [Y/Fe] and [Zr/Fe] (similar to RCBs) (Figure 3).
  V1920\,Cyg has more enhanced abundances of Y, Zr and the range in abundance
 variations is also very similar to RCBs (Pandey et al. 2004). 
 
  \begin{figure}[ht!]
\psfig{figure=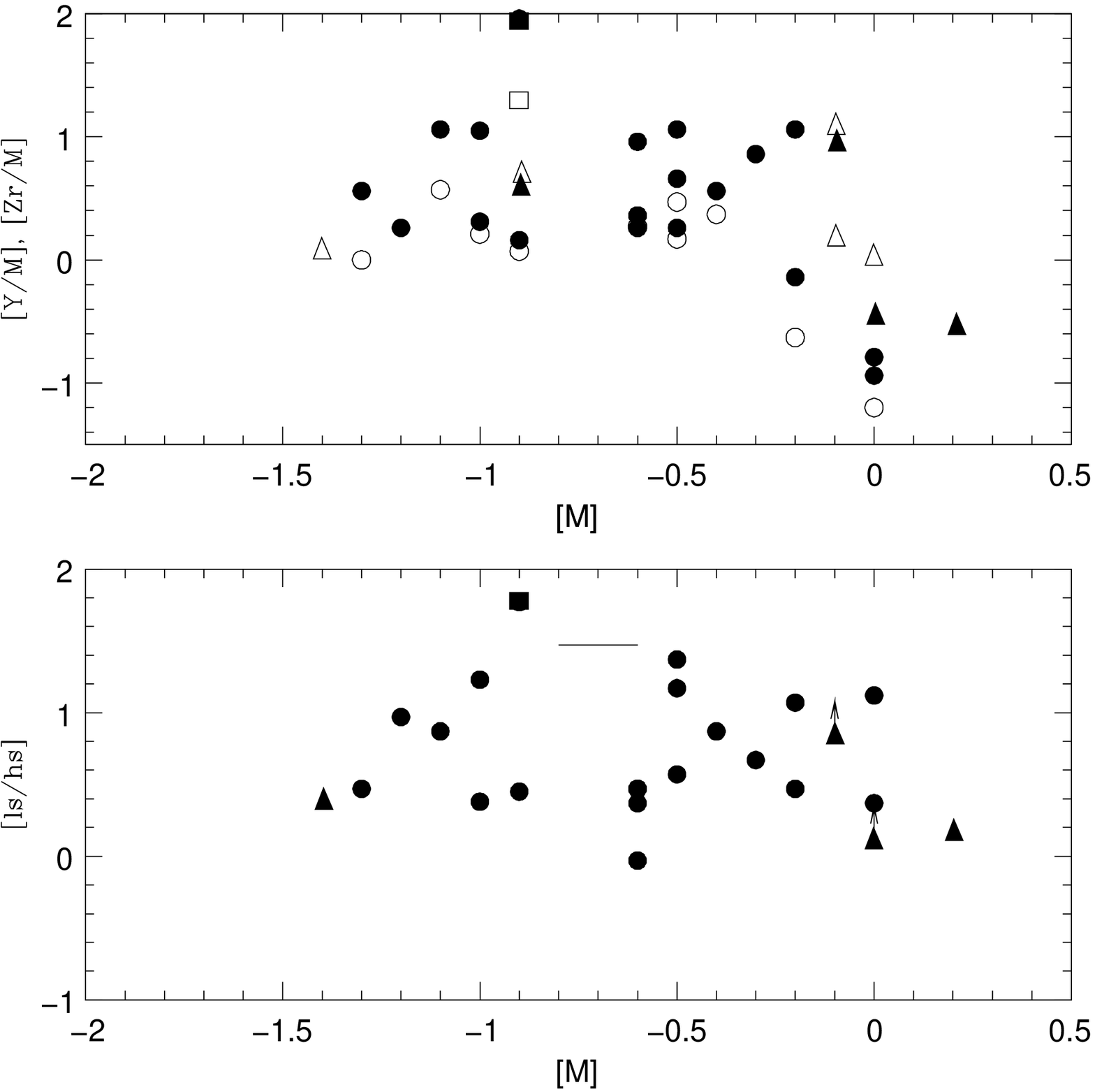,width=\hsize,height=14cm}
\caption{Distribution of light s-process elements Y (solids) and Zr (open symbols) 
with respect to metallicity parameter M for EHes (triangles) and RCBs (dots). The short line
refers to U Aqr, and the square refers to Sakurai's object.}
\end{figure}
 
 Although the
  abundances of heavy $s$-process elements could not be estimated, the upper
  limits of the abundances of Ce, Nd do demonstrate the lighter $s$-process
  elements are more enhanced even in EHe stars.
       It is generally acknowledged that $^{13}$C$(\alpha,n)^{16}$O is the main source of neutrons
  to run the $s$-processing in the He-burning shells of intermediate-mass AGB
  stars. Sufficient amounts of $^{13}$C are to be generated by slow mixing of
  protons into the $^{12}$C rich intershell regions to generate neutrons.
  The neutron irradiation occurs in radiative conditions. The heavier the
  neutron flux the greater is the abundance of heavies relative to light
  $s$-process elements.
     Busso et al. (2001) used the distribution of the ratio of heavy $s$-process
  celements (hs) to the light $s$-process (ls) elements  with respect to
  metallicity to characterize various parameters of neutron exposures during the third
  dredgeup phase in AGB stars eg. mass of $^{13}$C pocket in the inter shell regions.
  Reddy et al. (2002) showed that the variation
  of the [hs/ls] with respect to metallicity in post -AGB stars (that went through
  third dredgeup) is characterized by a model ST/1.5 of Busso et al. (2001).

  A plot of [Y/M] and [Zr/M] versus [M] for RCBs and EHes (figure 4 ) shows that the enhancements are positive and both show a similar range in their abundances.
   We compared the run of the ratio of [ls/hs] for RCBs and EHes with respect
  to the metallicity parameter [M] . The estimates for EHes are based on the
  upper limits for the heavy $s$-process elements and includes data from our
  ongoing analysis of the $HST$ UV spectra. Estimates of U\,Aqr and the born
  again giant, Sakurai's object (during May - Oct 1996) are also included
  for comparison. Both the groups RCBs and EHes blend together emphasizing
  the similarity in their ls/hs ratios.

\begin{figure}[ht!]
\psfig{figure=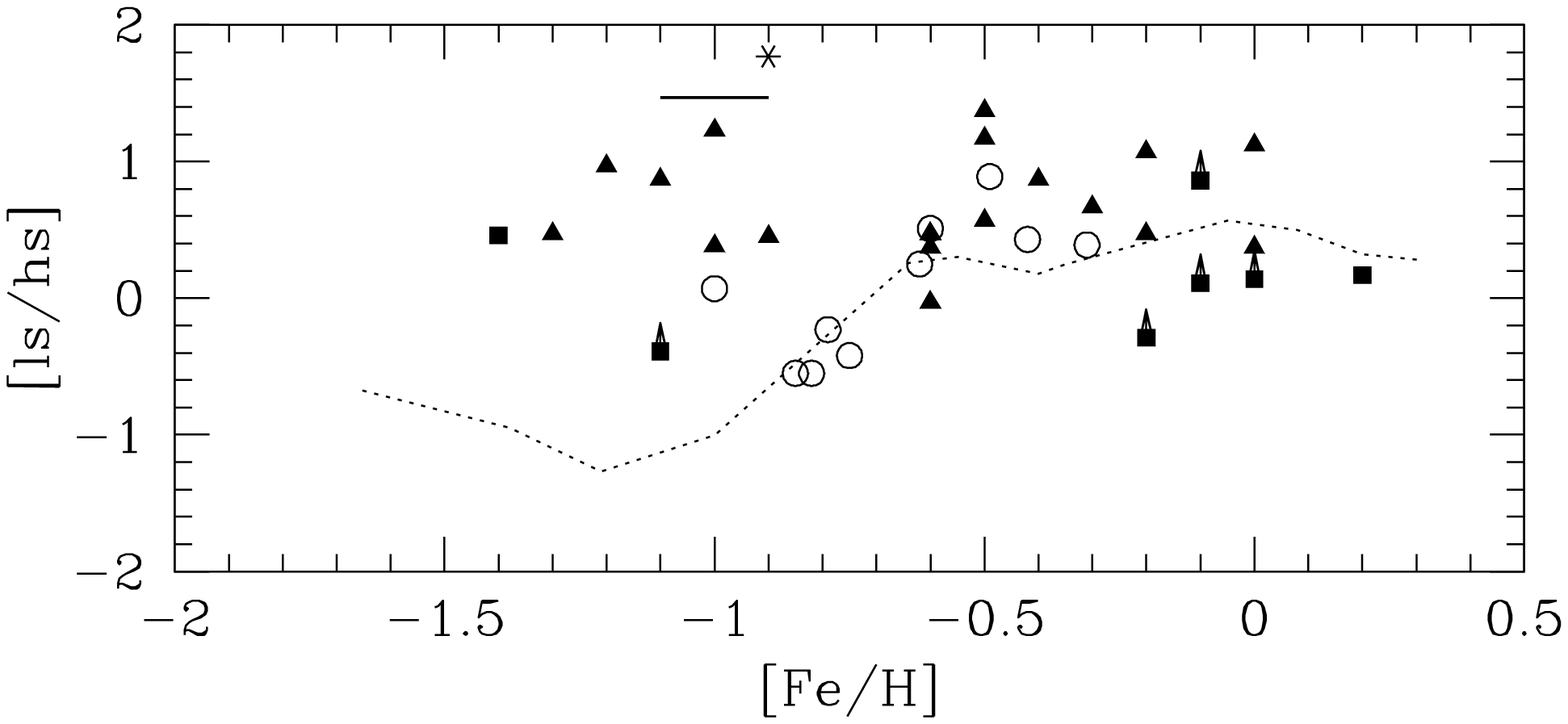,width=\hsize,height=6cm}
\caption{Ratio of light s-process elements to heavy s- process elements w.r.t 
metalicity. RCBs (triangles) and EHes (squares) are compared to post- AGB objects (open
circles) from Reddy et al.(2002). The dotted line refers to the model ST/1.5.}
\end{figure}

   Figure 5  shows a comparison of the
  trend of [ls/hs] in RCB and EHe stars along with that shown by post- AGB
  stars (Reddy et al. 2002) and the Busso et al's (2001) model ST/1.5.
  It is obvious that the trend of [ls/hs] with respect to metallicity
  of RCB and EHes is quite different from that shown by post-AGB stars.
  It also, probably, suggests the $s$-processing in RCBs and EHes is not a result
  of the third dredgeup and could have happened when the stars passed through
  a second AGB phase (presumably).

\section{Minority RCBs and Sakurai's object (V4334\,Sgr)}

      Are the minority RCBs born-again giants? The similarity of the abundance
  pattern of V854\,Cen with Sakurai's object, a clear example of born-again
  giant (final flash object), has been pointed out by Asplund et al. (1998).
  It is expected that the final flash objects show a C/He ratio much greater than
  1\%, say about 10 to 30\%. The C/He value of 10\% has been estimated for
  Sakurai's object in 1996 May to October (Asplund et al. 1998). Asplund et al.
  infer a C/He of 10\% for V854\,Cen. Asplund et al. (1998, 2000) also state that
  there are indications to suggest that the minority objects V3795\,Sgr and VZ\,Sgr have
  higher C/He values greater than 1\% -may even be 10\%. 
  
   \begin{figure}[ht!]
\psfig{figure=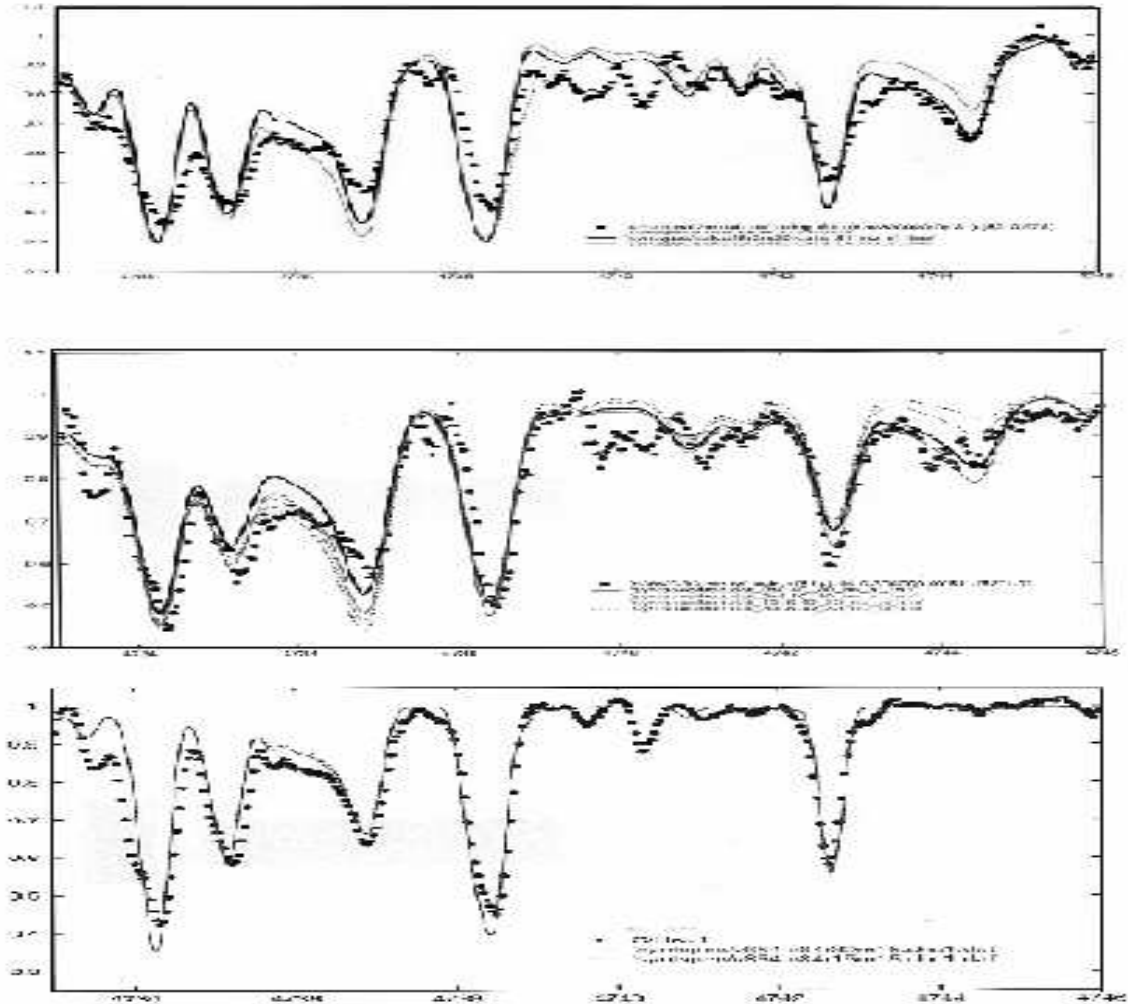,width=\hsize,height=14cm}
\caption{Synthesis of the spectral region of $^{12}$C$^{12}$C and $^{12}$C$^{13}$C bands in Sakurai's object (top), V CrA (middle) and V854 Cen (bottom). The $^{12}$C /$^{13}$C of 3 - 5, 4 - 10 and $>$ 30 is estimated for Sakurai's object, V CrA and V854 Cen,
respectively.} 
\end{figure}
  
      The abundance patterns of the minority stars suggested more diversity
  and range, in particular light elements like H, N, O and some  $s$-processed
  elements. Particularly V\,CrA showed differences relative to  other members.
  However, revised estimates from a recent analysis of higher resolution spectra
  (Rao \& Lambert 2004) suggests similar abundances for N, O, etc., as for the
   rest of the members. The abundances are more uniform in the minority group as well.
  The abundance pattern (i.e., log X/Fe) of V CrA is more similar to that of Sakurai's object  during May - October 1996 (Asplund et al. 1997, 1999) for most elements
  except Mg, Si, S, and Ca. The $s$-process elements in V CrA are also not enhanced as
  much as in Sakurai's object.

     One of the expectations of the final flash scenario is the presence of $^{13}$C, 
  and a low value of
  $^{12}$C/$^{13}$C ratio. It is expected that surface protons are mixed into the
  intershell region trigering CN cycle and converting $^{12}$C to $^{14}$N and $^{13}$C depending
  on the proton supply (Renzini 1990). In fact, Sakurai's object showed
 a $^{12}$C/$^{13}$C value of 4 although the other final flash objects like FG\,Sge (Gonzalez
 et al. 1998) and V605\,Aql (which showed a hydrogen deficient carbon star
  spectrum at maximum light -Lundmark 1919, Clayton \& De Marco 1997) did not
  show the presence of $^{12}$C$^{13}$C bands. Most  RCBs have been shown to have a high value
  of the $^{12}$C/$^{13}$C ratio. Keenan \& Barnbaum (1997) detected  the $^{12}$C$^{13}$C band at 4744\AA\ in
  the cool peculiar RCB variable DY\,Per.
   It was a surprise to find $^{12}$C$^{13}$C bands in the minority
  star V\,CrA (Rao \& Lambert 2004). We have synthesized the $^{12}$C$^{12}$C and $^{12}$C$^{13}$C
  bands of the Swan system (1,0) in the three stars; Sakurai's object in
  Oct 1996, V854\,Cen and V\,CrA (Figure 6) to match the observations.
  We used the line list and physical parameters obtained by Asplund et al.
  (1997) for Sakurai's object. The estimated ratio of $^{12}$C/$^{13}$C ranged between
  3 $-$ 5  for Sakurai's object (Asplund et al. 1997), has an upper limit of 30
  for V854\,Cen and a value 4 $-$ 10 for V\,CrA , thus displaying a similarity with Sakurai's object. The presence of $^{13}$C in the atmosphere of
  a RCB star does support the suggestion $^{13}$C$(\alpha,n)^{16}$O is the neutron
  source. It is possible that the evolutionary path for all minority RCBs
  (may even EHes) is through final flash.

   Discovery of a post -AGB hydrogen deficient star
  in globular cluster M5 (Dixon et al 2004) is an exciting new development
  which could pin down an age (and possible mass) to the progenitor.

    In summary, it now appears that at least some EHe stars show enhanced
  abundances of light $s$-process elements, e.g., Y, Zr as well as a ls/hs ratio
  similar to RCBs. The variation of the ls/hs ratio with decreasing metallicity
  suggests that $s$-processing in RCBs and EHes is not similar to that
  experienced by post -AGB (and AGB) stars (i.e., ST/1.5 model of Busso et al. 2001).
  The abundance ratios suggest a single exposure of $\tau_{o}$ of 0.1 to 0.2 mb$^{-1}$.
  It is likely that this episode of $s$-process element production might have
  occured when the stars were passing through AGB phase for a second time.
  The similarity in the abundance patterns of majority RCBs and majority EHes
  including the $s$-process elements and the presence of enhanced abundances of
  Ne and Mg in EHes does suggest that EHe phase might be later in evolution
  to that of RCBs. The minority RCBs seems to be a more coherent group in
  abundance distribution than earlier estimates indicated.
  Minority EHes and RCBs show a very similar abundance pattern, except for
  Si, Sc, and Ca, (elements that could be tied up in dust).
  RCBs have IR excesses and dust production episodes. The discovery of low
  $^{12}$C/$^{13}$C ratio (4 $-$ 10) in the minority RCB, V\,CrA does provide long awaited
  evidence for the mixing of surface protons to the intershell region and
  subsequent production of neutrons by $^{13}$C$(\alpha,n)^{16}$O, similar to Sakurai's
  object. The similarity of abundance patterns of V\,CrA and V854\,Cen to that
  displayed by Sakurai's object in 1996 Oct might encourage the suggestion
  that all minority RCBs are formed through final flash.

\section{Acknowledgements}

     I would like to thank my collaborators David Lambert, Gajendra Pandey,
  Simon Jeffery for letting me use some results before publication.
  I would also like to thank Martin Asplund for supplying me atmospheric models
  and line lists for $^{12}$C$^{13}$C bands. I would like to express my thanks to 
  David Yong,
  Eswar Reddy and Gajendra Pandey for preparing the figures and other help.
  I would also like express my appreciation to the organisers of the conference for
  their generous hospitality in Austin.

\section{References}

\begin{quote}
\verb"Alcock, C., Allsman, R. A., Alves, D. R., et al. 2001, ApJ,"\\
\verb"554, 298"

\verb"Asplund, M., Gustafsson, B., Kiselman, D., Eriksson, K."\\
\verb"1997a, A&A, 318, 521"

\verb"Asplund, M., Gustafsson, B., Lambert, D. L., Rao, N. K."\\
\verb"1997b, A&A, 321, L17"

\verb"Asplund, M., Gustafsson, B., Rao, N. K., Lambert, D. L."\\
\verb"1998, A&A, 332, 651"

\verb"Asplund, M., Lambert, D. L., Kipper, T., Pollacco, D., "\\
\verb"Shetron, M. D. 1999, A&A, 343, 507"

\verb"Asplund, M., Gustafsson, B., Lambert, D. L., Rao, N. K.,"\\
\verb"2000, A&A, 353, 287"

\verb"Bond, H. E., Luck, R. E., Newman, M. J. 1979, ApJ, 233,"\\
\verb"205"

\verb"Busso, M., Gallino, R., Lambert, D. L., Travaglio, C.,"\\
\verb"Smith, V. V. 2001, ApJ, 557, 802"

\verb"Clayton, G. C., de Marco, O. 1997, AJ, 114, 2679"

\verb"Cottrell, P. L., Lawson, W. A. 1998, PASA, 15, 179"

\verb"Dixon, W. V., Brown, T. M., Landsman, W. B. 2004, ApJ,"\\
\verb"600, L43"

\verb"Duerbeck, H. W., Liller, W., Sterken, C., et al. 2000,"\\
\verb"AJ, 119, 2360"

\verb"Gonzalez, G., Lambert, D. L., Wallerstein, G.,"\\
\verb"Rao, N. K., Smith, V. V., McCarthy, J. K. 1998, ApJS,"\\
\verb"114, 132"

\verb"Heber, U. 1986, in Hydrogen Deficient Stars,"\\
\verb"IAU Coll. 87, ed. K. Hunger, D. Schonberner & "\\
\verb"N. K. Rao, Reidel, Dordrecht, 32"

\verb"Iben, I. Jr., Tutukov, A. V. 1984, ApJS, 55, 335"

\verb"Iben, I, Jr., Tutukov, A. V., Yungelson, L. R. 1996,"\\
\verb"ApJ, 456, 750"

\verb"Iben, I. Jr., Kaler, J. B., Truran, J. W., Renzini, A."\\
\verb"1983, ApJ, 264, 605"

\verb"Jeffery, C. S. 1996, in ASP Conf. Ser. Vol. 96,"\\
\verb"Hydrogen Deficient Stars, ed. C. S. Jeffery & U. Heber, 152"

\verb"Jeffery, C. S., Heber, U. 1992, A&A, 260, 133"

\verb"Jeffery, C. S., Heber, U. 1993, A&A, 270, 167"

\verb"Jeffery, C. S., Drilling, J. S., Heber, U. 1987, MNRAS,"\\
\verb"226, 317"

\verb"Jeffery, C. S., Woolf, V. M., Pollacco, D. L. 2001, A&A,"\\
\verb"376, 497"

\verb"Keenan, P. C., Barnbaum, C. 1997, PASP, 109, 969"

\verb"Lambert, D. L., Rao, N. K. 1994, J. Astrophys. Astron.,"
\verb"15, 47"

\verb"Lundmark, K. 1921, PASP, 33, 814"

\verb"Morgan, D. H., Hatzidimitriou, D., Cannon, R. D., Croke,"\\
\verb"B. F. 2003, MNRAS, 344, 325"

\verb"Pandey, G., Rao, N. K., Lambert, D. L., Jeffery, C. S.,"\\
\verb"Asplund, M. 2001, MNRAS, 324, 937"

\verb"Pandey, G., Lambert, D. L., Rao, N. K., Jeffery, C. S."\\
\verb"2004a, ApJ, 602, L113"

\verb"Pandey, G., Lambert, D. L., Rao, N. K., Gustafsson, B.,"\\
\verb"Ryde, N., Yong, D. 2004b, MNRAS, 353, 143"

\verb"Rao, N. K., Lambert, D. L. 1996, in ASP Conf. Ser. Vol. 96,"\\
\verb"Hydrogen Deficient Stars, ed. C. S. Jeffery & U. Heber, 43"

\verb"Rao, N. K., Lambert, D. L. 2003, PASP, 115, 1304"

\verb"Rao, N. K., Lambert, D. L. 2004, (in preparation)"

\verb"Renzini,A. 1990, in ASP Conf. Ser, Vol. 11, Confrontation"\\
\verb"between Stellar Pulsation and Evolution, ed. C. Cacciari,"\\
\verb"G. Clementini, 549"

\verb"Reddy, B. E., Lambert, D. L., Gonzalez, G., Yong, D. 2002,"\\
\verb"ApJ, 564, 482"

\verb"Saio, H., Jeffery, C. S. 2000, MNRAS, 313, 671"

\verb"Saio, H., Jeffery, C.S. 2002, MNRAS, 333, 121"

\verb"Vanture, A. D., Zucker, D., Wallerstein, G. 1999, ApJ, 514,"\\
\verb"932"

\verb"Webbink, R. F. 1984, ApJ, 277, 355"
\end{quote}

\section{ }
J. C. Wheeler
: Can you comment on mass measurements for
the RCB and EHe stars?

K.Rao: Simon Jeffery and Vincent Woolf have tried
to obtain estimates for two pulsating EHe stars V652
Her, LSS 3184.  They obtain $\sim$ 0.5 $M_{\odot}$ for the mass.
LSS 3184 is carbon rich.  If a CO white dwarf has to get
into action, this mass estimate is little low.

J. Lattanzio
: Can you determine the Mg isotopes for your
Mg-rich stars?

K.Rao: No. We have to find some molecule containing
Mg to estimate isotope ratios. Since H is weak, MgH is
not present, even in cool stars. It is a difficult problem.

J. Cohen
: You and some previous speakers have
spoken about abundance peculiarities. We need to
make sure that the abundance peculiarities are real and
not the result of the physical phenomena of
gravitational settling and radiative levitation such as
seen on the blue HB of globular clusters.

K.Rao: These stars are supergiants with low effective
gravities, with rotation, and  appreciable
microturbulance.  As such it is unlikely to expect
gravitational settling to operate.  Moreover, the
abundance pattern is not similar to that seen in blue
horizontal branch stars.

\end{document}